\documentstyle[preprint,aps,epsf]{revtex}
\draft
\tightenlines
\newcommand{\beq}{\begin{equation}}
\newcommand{\eeq}{\end{equation}}

\begin{document}

\title{ Quasiparticles in the Superconducting State of High-T$_c$
Metals }

\author{M.Ya. Amusia$^{a,b}$ and V.R.
Shaginyan$^{a,c}$ \footnote{E--mail: vrshag@thd.pnpi.spb.ru}}

\address{ $^{a\,}$The Racah Institute of Physics,
the Hebrew University, Jerusalem 91904, Israel;\\
$^{b\,}$A.F. Ioffe Physical-Technical Institute, 194021 St.
Petersburg, Russia;\\
$^{c\,}$Petersburg Nuclear Physics Institute, Gatchina, 188300,
Russia}

\maketitle

\begin{abstract}

We consider the behavior of quasiparticles in the superconducting
state of high-$T_c$ metals within the framework of the theory of
superconducting state  based on the fermion condensation quantum
phase transition. We show that the behavior coincides with the
behavior of Bogoliubov quasiparticles, whereas the maximum value
of the superconducting gap and other exotic properties are determined
by the presence of the fermion condensate. If at low temperatures the
normal state is recovered by the application of a magnetic field
suppressing the superconductivity, the induced state can be viewed as
Landau Fermi liquid. These observations  are in good agreement with
recent experimental facts.

\end{abstract}


The Landau Fermi Liquid (LFL) theory explains the major part of
the low-temperature properties of Fermi liquids \cite{lan}. The
LFL theory has demonstrated that the low-energy elementary
excitations of a Fermi liquid look like the spectrum of an ideal
Fermi gas and can be described in terms of Landau quasiparticles
(LQ) with an effective mass $M^*$, charge $e$ and spin $1/2$. As
well, the LFL theory gives theoretical grounds for the BCS
(Bardeen, Cooper, and Schrieffer) theory \cite{bcs} of
conventional superconductivity which accounts for many of
fundamental properties of superconductors. In turn, the BCS theory
is based on the notion of quasiparticles which represent
elementary excitations of superconducting electron liquid and are
called Bogoliubov quasiparticles (BQ). In the case of high-$T_c$
metals, when the understanding of their striking behavior remains
among the main problems of the condensed matter physics, a number
of primary ideas of the LFL theory and BCS theory has been called
in question. Therefore, there exists a fundamental question about
whether or not a theory of high-$T_c$ metals can be developed in
terms of LQ and BQ.

It was reported recently that the full energy dispersion of
single-particle excitations and the corresponding coherence
factors as a function of momentum were measured on high-$T_c$
cuprate (Bi$_2$Sr$_2$Ca$_2$Cu$_3$O$_{10+\delta}$, $T_c$=108 K) by
using high-resolution angle-resolved photoemission spectroscopy
\cite{mat}. All the observed features qualitatively and
quantitatively agree with the behavior of BQ predicted from BCS
theory. This observation suggests that the superconducting state
of high-$T_c$ cuprate is BCS-like and implies the basic validity
of BCS formalism in describing the superconducting state
\cite{mat}. On the other hand, such properties as the pairing
mechanism, the maximum value of the superconducting gap
$\Delta_1$, the high density of states, and other exotic
properties are beyond BCS theory.

Striking experimental facts on the transport properties of the
normal state induced by applying a magnetic field greater than the
upper critical filed $B_c$ were obtained  in a hole doped cuprates
at overdoped concentration (Tl$_2$Ba$_2$CuO$_{6+\delta}$)
\cite{cyr} and at optimal doping concentration
(Bi$_2$Sr$_2$CuO$_{6+\delta}$) \cite{cyr1}. These data have
clearly shown that there are no any sizable violation of the
Wiedemann-Franz (WF) law. Measurements for strongly overdoped
non-superconducting La$_{1.7}$Sr$_{0.3}$CuO$_4$ have demonstrated
that the resistivity $\rho$ exhibits $T^2$ behavior,
$\rho=\rho_0+\Delta\rho$ with $\Delta\rho=AT^2$, and the WF law is
verified to hold perfectly \cite{nakam}. Since the validity of the
WF law is a robust signature of LFL, these experimental facts
demonstrate that the observed elementary excitations cannot be
distinguished from LQ. Thus these
experimental observations impose strong constraints for models
describing the electron liquid of the high-temperature
superconductors. For example, in the cases of a Luttinger liquid
\cite{kane}, spin-charge separation (see e.g. \cite{sen}), and in
some solutions of $t-J$ model \cite{hough} a violation of the WF
law was predicted.

In this Letter, we consider the superconducting state of
high-$T_c$ metals within the framework of the theory of
superconducting state  based on the fermion condensation quantum
phase transition (FCQPT) \cite{ks,ms,ms1}.
We show that the superconducting state is
BCS-like, the elementary excitations are BQ, and the primary ideas
of the LFL theory and BCS theory are valid. At temperatures $T\to
0$, the normal state recovered by the application of a magnetic
field larger than the critical field $B_c$ can be viewed as  LFL
induced by the magnetic field. In this state, the WF law is held
and the elementary excitations are LQ.

At $T<T_c$, the free energy $F$ of an electron liquid is given the
equation (see, e.g. \cite{til}) \beq F-\mu N=E_{gs}-\mu N-TS,\eeq
where $N$ is the number of particles, $S$ denotes the entropy,
and $\mu$ is the chemical potential.
The ground state energy $E_{gs}[\kappa({\bf p}),n({\bf p})]$ of an
electron liquid is a functional of the order parameter of the
superconducting state $\kappa({\bf p})$ and of the quasiparticle
occupation numbers $n({\bf p})$. Here we assume that the electron
system is two-dimensional, while all results can be transported to
the case of three-dimensional system. This energy is determined by
the known equation of the weak-coupling theory of
superconductivity \beq E_{gs}\ =\ E[n({\bf p})]+\int
\lambda_0V({\bf p}_1,{\bf p}_2) \kappa({\bf p}_1) \kappa^*({\bf
p}_2) \frac{d{\bf p}_1d{\bf p}_2}{(2\pi)^4}\ . \eeq Here $E[n({\bf
p})]$ is the Landau functional determining the ground-state energy
of a normal Fermi liquid. The quasiparticle occupation numbers
\beq n({\bf p})=v^2({\bf p})(1-f({\bf p}))+u^2({\bf p})f({\bf p})
,\eeq and \beq\kappa({\bf p})=v({\bf p})u({\bf p})(1-2f({\bf
p})),\eeq where the coherence factors $v({\bf p})$ and $u({\bf p})$
are obeyed the normalization condition \beq v^2({\bf p})+u^2({\bf
p})=1.\eeq The distribution function $f({\bf p})$ of BQ defines
the entropy \beq S=-2\int\left[f({\bf p})\ln f({\bf p})+(1-f({\bf
p}))\ln(1-f({\bf p}))\right]\frac{d{\bf p}}{4\pi^2}.\eeq

We assume that the pairing interaction $\lambda_0V({\bf p}_1,{\bf
p}_2)$ is weak and produced, for instance, by electron-phonon
interaction. Minimizing $F$ with respect to $\kappa({\bf p})$ and
using the definition $\Delta({\bf p})=-\delta F/\kappa({\bf p})$,
we obtain the equation connecting the single-particle energy
$\varepsilon({\bf p})$ to the superconducting gap $\Delta({\bf
p})$, \beq \varepsilon({\bf p})-\mu\ =\ \Delta({\bf p})
\frac{1-2v^2({\bf p})} {2v({\bf p})u({\bf p})}. \eeq The
single-particle energy $\varepsilon({\bf p})$ is determined by the
Landau equation \beq \varepsilon({\bf p})= \frac{\delta E[n({\bf
p})]}{\delta n({\bf p})}. \eeq Note that $E[n({\bf p})]$,
$\varepsilon[n({\bf p})]$, and the Landau amplitude \beq F_L({\bf
p},{\bf p}_1)=\frac{\delta E^2[n({\bf p})]}{\delta n({\bf
p})\delta({\bf p}_1)}\eeq implicitly depend on the density $x$
which defines the strength of $F_L$. Minimizing $F$ with respect
to $f({\bf p})$ and after some algebra, we obtain the equation for
the superconducting gap $\Delta({\bf p})$ \beq \Delta({\bf p})=
-\frac{1}{2}\int\lambda_0 V({\bf p},{\bf p}_1) \frac{\Delta({\bf
p}_1)}{E({\bf p}_1)}(1-2f({\bf p})) \frac{d{\bf p}_1}{4\pi^2}.
\eeq Here the excitation energy $E({\bf p})$ of BQ is given by
\beq E({\bf p})=\frac{\delta (E-\mu N)}{\delta f({\bf p})}=
\sqrt{(\varepsilon({\bf p})-\mu)^2+\Delta^2({\bf p})}. \eeq The
coherence factors  $v({\bf p})$, $u({\bf p})$, and
the distribution
function $f({\bf p})$ are given by the ordinary relations \beq
v^2({\bf p})=\frac{1}{2}\left(1-\frac{\varepsilon({\bf
p})-\mu}{E({\bf p})}\right),\,\,\, u^2({\bf
p})=\frac{1}{2}\left(1+\frac{\varepsilon({\bf p})-\mu}{E({\bf
p})}\right),\eeq \beq f({\bf p})=\frac{1}{1+\exp(E({\bf
p})/T)}.\eeq Equations (7-13) are the conventional equations of
the BCS theory \cite{bcs,til}, determining the superconducting state
with BQ and the maximum value of the superconducting gap
$\Delta_1\sim 10^{-3}\varepsilon_F$ provided that one assumes that
system in question has not undergone FCQPT.

Now we turn to a consideration of superconducting electron liquid
with the fermion condensate (FC) which takes place after the
FCQPT point. If $\lambda_0\to 0$, then the maximum
value of the superconducting gap $\Delta_1\to 0$, as well as the
critical temperature $T_c\to0$, and Eq. (7) reduces to the equation
\cite{ks,ms,ams} \beq \varepsilon({\bf p})-\mu\ =\ 0,\quad \mbox{
if}\quad 0<n({\bf p})<1;\: p_i\leq p\leq p_f\ . \eeq At $T\to0$,
Eq. (14) defines a new state of electron liquid with
FC \cite{ks,vol}
which is characterized by a flat part of the
spectrum in the $(p_f-p_i)$ region and has a strong impact on
the system's properties up to temperature $T_f$
\cite{ks,ms,duk}. Apparently, the
momenta $p_i$ and $p_f$ have to satisfy $p_i<p_F<p_f$, where $p_F$
is the Fermi momentum. When the Landau amplitude
$F_L(p=p_F,p_1=p_F)$  as a function of the density $x$ is
sufficiently small, the flat part vanishes, and at $T\to0$ Eq. (14)
has the only trivial solution $\varepsilon(p=p_F)=\mu$, and the
quasiparticle occupation numbers are given by the step function,
$n({\bf p})= \theta(p_F-p)$ \cite{ks}. At some critical density
$x=x_{FC}$ the amplitude becomes strong enough so that Eq. (14)
possesses the solution corresponding to a formation of the flat
part of spectrum, that is FC is created \cite{ksz,shag}. Note,
that a formation of the flat part of the spectrum has been
recently confirmed in Ref. \cite{irk}.

Now we can study the relationships between the state defined by
Eq. (14) and the superconductivity. At $T\to0$, Eq. (14) defines a
particular state of a Fermi liquid with FC, for which the modulus
of the order parameter $|\kappa({\bf p})|$ has finite values in
the $(p_f-p_i)$ region, whereas  $\Delta_1\to 0$ in this region.
Observe that $f({\bf p},T\to0)\to0$, and it follows from Eqs. (3)
and (4) that if $0<n({\bf p})<1$ then $|\kappa({\bf p})|\neq 0$ in
the region $(p_f-p_i)$ . Such a state can be considered as
superconducting, with an infinitely small value of $\Delta_1$, so
that the entropy of this state is equal to zero. It is obvious
that this state being driven by the quantum phase transition
disappears at $T>0$ \cite{ms}. Any quantum phase transition, which
takes place at temperature $T=0$, is determined by a control
parameter other then temperature, for example, by pressure, by
magnetic field, or by the density of mobile charge carriers $x$.
The quantum phase transition occurs at a quantum critical point.
At some density $x\to x_{FC}$, when the Landau amplitude $F_L$
becomes sufficiently weak, and $p_i\to p_F\to p_f$, Eq. (14)
determines the critical density $x_{FC}$ at which
FCQPT takes place leading
to the formation of FC \cite{ks,ms}. It follows from Eq. (14) that
the system becomes divided into two quasiparticle subsystems: the
first subsystem in the $(p_f-p_i)$ range is characterized  by the
quasiparticles with the effective mass $M^*_{FC}\propto
1/\Delta_1$, while the second one is occupied by quasiparticles
with finite mass $M^*_L$ and momenta $p<p_i$. The density of
states near the Fermi level tends to infinity, $N(0)\propto
M^*_{FC}\propto 1/\Delta_1$ \cite{ms}.

If $\lambda_0\neq 0$, then $\Delta_1$ becomes finite. It is seen
from Eq. (10) that the superconducting gap depends on the
single-particle spectrum $\varepsilon({\bf p})$. On the other
hand, it follows from Eq. (7) that $\varepsilon({\bf p})$
depends on $\Delta({\bf p})$ provided that at $\Delta_1\to 0$ Eq.
(14) has the solution determining the existence of FC. Let us
assume that $\lambda_0$ is small so that the particle-particle
interaction $\lambda_0 V({\bf p},{\bf p}_1)$ can only lead to a
small perturbation of the order parameter $\kappa({\bf p})$
determined by Eq. (14). Upon differentiation both parts of Eq. (7)
with respect to the momentum $p$, we obtain that the effective
mass $M^*_{FC}=d\varepsilon(p)/dp_{\,|p=p_F}$ becomes finite
\cite{ms} \beq M^*_{FC}\sim p_F\frac{p_f-p_i}{2\Delta_1}. \eeq It
follows from Eq. (15) that the effective mass and the density of
states $N(0)\propto M^*_{FC}\propto 1/\Delta_1$ are finite and
constant at $T<T_c$ \cite{ms,ams}. As a result, we are led to the
conclusion that in contrast to the conventional theory of
superconductivity the single-particle spectrum $\varepsilon({\bf
p})$ strongly depends on the superconducting gap and we have to
solve Eqs. (8) and (10) in a self-consistent way. On the other
hand, let us assume that Eqs. (8) and (10) are solved, and
the effective mass $M^*_{FC}$ is determined. Now one
can fix the dispersion $\varepsilon({\bf p})$ by choosing
the effective mass $M^*$ of system in question equal to
$M^*_{FC}$ and then solve Eq. (10) as it is done in the case of
the conventional theory of superconductivity \cite{bcs}. As a
result, one observes that the superconducting state is
characterized by BQ with the dispersion
given by Eq. (11), the coherence factors  $v$, $u$
are given by Eq. (12), and the normalization condition
(5) is held. We are lead to
the conclusion that the observed features agree with the behavior
of BQ predicted from BCS theory. This observation suggests that
the superconducting state with FC is BCS-like and implies the
basic validity of BCS formalism in describing the superconducting
state. It is exactly the case that was observed experimentally in
high-$T_c$ cuprate Bi$_2$Sr$_2$Ca$_2$Cu$_3$O$_{10+\delta}$
\cite{mat}.

Consider other differences between the conventional
superconducting state and the superconducting state with FC.
We consider the case when $T_c\ll T_f$.  This means that the
order parameter $\kappa({\bf p})$ is
slightly perturbed by the pairing
interaction because the particle-particle
interaction $\lambda_0 V$ is small comparatively to the Landau
amplitude $F_L$ and the order parameter $\kappa({\bf p})$ is
governed mainly by $F_L$ \cite{ks}. We can solve Eq. (10)
analytically taking the Bardeen-Cooper-Schrieffer approximation for
the particle-particle interaction:  $\lambda_0V({\bf p},{\bf
p}_1)=-\lambda_0$ if $|\varepsilon({\bf p})-\mu|\leq \omega_D$, i.e.
the interaction is zero outside this region, with $\omega_D$ being
the characteristic phonon energy. As a result, the maximum value of
the superconducting gap is given by \cite{ams}
\beq \Delta_1\simeq \frac{\lambda_0 p_F(p_f-p_F)}{2\pi}
\ln\left(1+\sqrt2\right)
\simeq 2\beta\varepsilon_F
\frac{p_f-p_F}{p_F}\ln\left(1+\sqrt2\right).  \eeq Here, the Fermi
energy $\varepsilon_F=p_F^2/2M^*_L$, and the dimensionless coupling
constant $\beta$ is given by the relation $\beta=\lambda_0
M^*_L/2\pi$. Taking the usual values of $\beta$ as $\beta\simeq 0.3$,
and assuming $(p_f-p_F)/p_F\simeq 0.2$, we get from Eq. (16) a large
value of $\Delta_1\sim 0.1\varepsilon_F$, while for normal metals
one has $\Delta_1\sim 10^{-3}\varepsilon_F$. Now we determine the
energy scale $E_0$ which defines the region occupied by
quasiparticles with the effective mass $M^*_{FC}$ \beq E_0=
\varepsilon({\bf p}_f)-\varepsilon({\bf p}_i) \simeq 2
\frac{(p_f-p_F)p_F}{M^*_{FC}}\ \simeq\ 2\Delta_1. \eeq

We have returned back to the Landau Fermi liquid theory since high
energy degrees of freedom are eliminated and the quasiparticles
are introduced. The only difference between LFL, which serves as a
basis when constructing the superconducting state, and Fermi
liquid after FCQPT is that we have to expand the number of
relevant low energy degrees of freedom by introducing a new type of
quasiparticles with the effective mass $M^*_{FC}$ given by Eq.
(15) and the energy scale $E_0$ given by Eq. (17). Therefore, the
dispersion $\varepsilon({\bf p})$ is characterized by two
effective masses $M^*_L$ and $M^*_{FC}$ and by the scale $E_0$,
which define the low temperature properties including the line
shape of quasiparticle excitations \cite{ms,ams}, while the
dispersion of BQ is given by Eq. (11). We note that both the
effective mass $M^*_{FC}$ and the scale $E_0$ are temperature
independent at $T<T_c$, where $T_c$ is the critical temperature of
the superconducting phase transition \cite{ams}. Obviously, we
cannot directly relate these new LFL quasiparticle excitations
with the quasiparticle excitations of an ideal Fermi gas because
the system in question has undergone FCQPT. Nonetheless, the main
basis of the Landau Fermi liquid theory survives FCQPT: the low
energy excitations of a strongly correlated liquid with FC are
quasiparticles.

As it was shown above, properties of these new quasiparticles are
closely related to the properties of the superconducting state. We
may say that the quasiparticle system in the range $(p_f-p_i)$
becomes very ``soft'' and is to be considered as a strongly
correlated liquid. On the other hand, the system's properties and
dynamics are dominated by a strong collective effect having its
origin in FCQPT and determined by the macroscopic number of
quasiparticles in the range $(p_f-p_i)$. Such a system cannot be
perturbed by the scattering of individual quasiparticles and has
features of a ``quantum protectorate" \cite{ms,rlp,pa}.

At $T_c<T$, the order parameter $\kappa$ vanishes, and the
behavior of system in question can be viewed as the behavior of an
anomalous electron Fermi liquid, or strongly correlated liquid,
with the resistivity being a linear function of  temperature,
while the effective mass behaves as $M^*_{FC}\propto 1/T$
\cite{ms,duk}. Obviously, at this regime one observes strong
deviations from the LFL behavior and cannot expect the WF law is
held.

As any phase transition, FCQPT is related to the order parameter,
which induces a broken symmetry. As we have seen, the order
parameter is the superconducting order parameter $\kappa({\bf
p})$, while $\Delta_1$ being proportional to the coupling constant
(see Eq. (16)) can be small. Therefore, the existence of such a
state, that is electron liquid with FC, can be revealed
experimentally. Since the order parameter $\kappa({\bf p})$ is
suppressed by the critical magnetic field $B_c$, when $B_c^2\sim
\Delta_1^2$. If the coupling constant $\lambda_0\to 0$, the weak
critical magnetic field $B_c\to 0$ will destroy the state with FC
converting the strongly correlated Fermi liquid into LFL. In this
case the magnetic field plays a role of the control parameter
determining the effective mass \cite{pogsh}
\begin{equation}
M^*_{FC}\propto \frac{1}{\sqrt{B}}.
\end{equation}

Equation (18) shows that by applying a magnetic field $B$ the
system can be driven back into LFL with the effective mass
$M^*_{FC}$ which is finite and independent of the temperature.
This means that the low temperature properties depend on the
effective mass in accordance with the LFL theory. At $T>T^*$, the
system possesses the behavior of the strongly correlated liquid.
Here $T^*\propto\sqrt{B}$ is the temperature at which the
transition from LFL to the strongly correlated liquid takes place.
Such a behavior was observed experimentally in the heavy-electron
metal YbRh$_2$Si$_2$ \cite{gen}. If $\lambda_0$ is finite, the
critical field is also finite, and Eq. (18) is valid at $B>B_c$.
In that case, the effective mass $M^*_{FC}$ is finite and
temperature independent at $T<T_c$, and low temperature elementary
excitations of the system can be described in terms of LQ. Thus,
system is driven back to LFL and has the LFL behavior induced by the
magnetic field at least at $T<T_c$.  While the low energy elementary
excitations are characterized by $M^*_{FC}$ and cannot be
distinguished from LQ. As a result, at $T\to 0$,
the WF law is held in accordance with experimental facts
\cite{cyr,cyr1}.

The existence of FCQPT can also be revealed experimentally because
at densities $x>x_{FC}$, or beyond the FCQPT point, the system
should be LFL at sufficiently low temperatures \cite{shag}. Recent
experimental data have shown that this liquid exists in heavily
overdoped non-superconducting La$_{1.7}$Sr$_{0.3}$CuO$_4$
\cite{nakam}. It is remarkable that up to $T=55$ K the resistivity
exhibits the $T^2$ behavior and the WF law is verified to within
the experimental resolution \cite{nakam}.

In summary, we have shown that the superconducting state with FC is
characterized by BQ. The behavior of these BQ agrees with the
behavior of BQ predicted from BCS theory and suggests that the
superconducting state with FC is BCS-like and implies the basic
validity of BCS formalism in describing the superconducting state.
Although the maximum value
of the superconducting gap and other exotic properties are determined
by the presence of the fermion condensate. We
have also demonstrated that the low temperature transport properties
of high-$T_c$ metals observed in optimally doped and overdoped
cuprates by the application of a magnetic field higher than the
critical field can be explained within the framework of the
fermion condensation theory of high-$T_c$ superconductivity. The
quasiparticles are LQ and the WF law is held. The recent
experimental observations of BQ in the superconducting state and
verifications of the WF law in heavily overdoped, overdoped and
optimally doped cuprates clearly favor the existence of FC in
high-$T_c$ metals.

MYA is grateful to the Hebrew University Intramural fund of the
Hebrew University for financial support. VRS is grateful to Racah
Institute of Physics for the hospitality during his stay in
Jerusalem. This work was supported in part by the Russian
Foundation for Basic Research, project No 01-02-17189.

\end{document}